\documentclass[aps,prb,superscriptaddress,showpacs,natbib,twocolumn]{revtex4}
\usepackage[english]{babel}
\usepackage{color}
\usepackage{graphicx}
\usepackage{cancel}
\usepackage{bm}
\usepackage{amsmath}
\usepackage{amssymb}

\def  \bsig    {\mbox{\boldmath$\sigma$}}
\def  \bta     {\mbox{\boldmath$\tau$}}

\keywords{graphene, spin current, spin polarization, inverse Edelstein effect}

\begin{document}
\title{Thermoelectric and thermospin transport in a ballistic junction of graphene}
\author{M. Inglot}
\affiliation{Research and Development Centre for Photovoltaics, ML System Sp. z o.o. Rzesz\'ow ul.~Warszawska 50D, 35-230 Rzesz\'ow, Poland}


\author{V. K. Dugaev}
\affiliation{Department of Physics, Rzesz\'ow University of Technology,
al.~Powsta\'nc\'ow Warszawy 6, 35-959 Rzesz\'ow, Poland}
\affiliation{Departamento de F\'isica and CeFEMA, Instituto Superior T\'ecnico,
Universidade de Lisboa, Av. Rovisco Pais, 001-1049 Lisboa, Portugal}

\author{J. Barna\'s}
\affiliation{Faculty of Physics, Adam Mickiewicz University, ul. Umultowska 85,
61-614 Pozna\'n, Poland}
\affiliation{The Nano-Bio-Medical Centre, Umultowska 85, 61-614 Pozna\'n, Poland}

\begin{abstract}
We consider theoretically a wide graphene ribbon, that on both ends is attached to electronic reservoirs which generally have  different temperatures. The graphene ribbon is assumed to be deposited on a substrate, that leads to a spin-orbit coupling of Rashba type.  We calculate the thermally induced charge current in the ballistic transport regime as well as the thermoelectric voltage (Seebeck effect). Apart from this, we also consider thermally induced spin current and spin polarization of the graphene ribbon. The spin currents are shown to have generally two components; one parallel to the temperature  gradient and the other one perpendicular to this gradient. The latter corresponds to the spin current due to the spin Nernst effect. Additionally, we also consider the heat current between the reservoirs due to transfer of electrons.
\end{abstract}

\date{\today}
\pacs{72.25.Fe, 78.67.Wj, 81.05.ue, 85.75.-d}

\maketitle

\section{Introduction}

Low energy electronic states in graphene -- a two-dimensional hexagonal lattice of carbon atoms -- are usually described by the relativistic
Dirac model.\cite{castro09}  Unique transport properties of graphene,  especially the high electron velocity and tendency to avoid electron scattering due to the Klein effect, make graphene an excellent material for future applications in nanoelectronics.\cite{sonin09,katsnelson06}  These properties of graphene also facilitate practical realization of ballistic junctions with graphene.\cite{lee15,borunda13}  Indeed, such junctions have been extensively studied in recent years.\cite{titov06,grushina13,nguyen14} It has been shown, for instance, that junctions of two graphene parts corresponding to different Rashba coupling parameters exhibit interesting transport features.\cite{Yamakage2011} When the graphene is additionally magnetized, e.g. due to coupling to an insulating magnetic substrate, a large anisotropic magnetoresistance effect can be then observed.\cite{Rataj13}

Remarkably less theoretical and experimental work has been done up to now on thermoelectric properties of
graphene, though interest in these properties is growing recently.\cite{balandin2008,Wierzbowska1,Wierzbowska2,Wierzbicki13,foster09,wei09,yan09,
zuev09,ugarte11,wang11,sharapov12,alomar14} Theoretical works were focused mainly on the diffusion transport regime in graphene with impurities and other structural defects.
It has been shown, for instance, that at certain conditions the
Wiedemann-Franz law can be violated in graphene.\cite{Seol10}
Moreover, resonant scattering from impurities with short-range potential may lead to an enhanced
Seebeck coefficient, when the chemical potential is in the
neighborhood of the resonances.\cite{Dragoman,stauberPRB07,Inglot15,Xu2014}
Of particular interest are currently thermoelectric properties of graphene nanoribbons,
which may exhibit an enhanced thermoelectric efficiency.\cite{Zheng09,Zhang10,aksamijaPRB14,
sevincli10,huang11,chang12,liang12}
This efficiency may be additionally enhanced by certain structural defects like
antidots, for instance.\cite{Wierzbicki13}
In addition, there is currently a great interest in spin related thermoelectric phenomena in nanoscale systems, including also graphene nanostructures.
It has been shown, among others, that graphene nanoribbons with zigzag edges can exhibit not only conventional but also spin thermoelectricity.\cite{Wierzbicki13}
The latter corresponds to a spin voltage generated by a temperature gradient. Various physical phenomena associated with thermally induced spin and heat currents are also of current interest.\cite{chenPRB14,lindsayPRB14}

One may observe recently an increasing interest in the ballistic transport regime.
\cite{oltscher14,jaffres14}  This is due to the possibility of a long mean free path $\ell $ in
2D electron gas \cite{oltscher14}, where $\ell \simeq 3$~$\mu $m has been already
reached,\cite{mayorov11} and in graphene nanoribbons, where $\ell >10$~$\mu $m has been reported.\cite{baringhaus14}
The main objective of this paper is a theoretically description of thermoelectric and thermospin transport properties of
ballistic graphene junctions.
We calculate the thermoelectrically induced
charge and spin currents in a graphene ribbon of length $L$, which is  attached  to two electronic reservoirs. The ribbon length is
assumed to be  smaller than the corresponding electron mean free path $\ell $.
The main focus in the paper is on the spin effects due to spin-orbit interaction in graphene. Since the intrinsic spin-orbit coupling in graphene is very small, it is neglected here. In turn, the Rashba spin-orbit
coupling related to the influence of a substrate can be relatively strong, and therefore it is included in our considerations.
In the presence of Rashba spin-orbit coupling in graphene, temperature gradient can generate not only the spin current
(spin Nernst effect) but also a spin density. It is well known, that spin current can be then generated also by an electric field (spin Hall effect). Similarly, the spin polarization may be induced by the electric field as well.\cite{Dyrdal14}

In section 2 we describe the theoretical model. Numerical results on the thermally induced charge and heat currents are
presented and discussed in section 3. In turn, thermally induced spin current and spin polarization of graphene is considered in section 4.
Summary and final conclusions are in section 5.

\section{Model}

We assume the relativistic Hamiltonian for electrons in graphene with Rashba spin-orbit coupling,
\begin{eqnarray}
\label{1}
H_{\bf k}=\hbar v_F\bta \cdot {\bf k}+\alpha (\sigma _x\tau _y-\sigma _y\tau _x) ,
\end{eqnarray}
where $v_F$ is the electron velocity in graphene, ${\bf{k}}=(k_x,k_y)$ is a two-dimensional wavevector, $\alpha $ is the Rashba
coupling parameter, while $\bta $ and $\bsig $ are the vectors of Pauli matrices
defined in the sublattice and spin spaces, respectively.
Hamiltonian (1) describes low-energy electron states in the vicinity
of the Dirac point $K$ of the corresponding  Brillouin zone. Hamiltonian for the second non-equivalent Dirac
point, $K'$, can be obtained from Eq.(1) by reversing sign of the wavevector component
$k_x$.
\begin{figure}[h]
\label{fig1}
\includegraphics[width=.7\columnwidth]{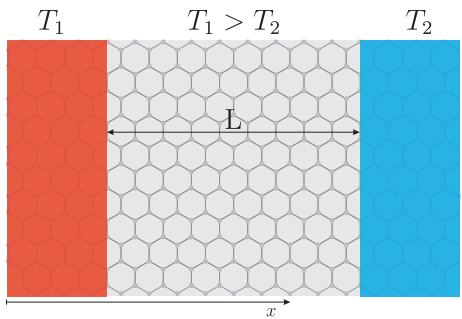}
\caption{Schematic of a  ballistic junction consisting of  a graphene ribbon of length $L$ and two 2D electronic reservoirs.
The reservoirs have generally different temperatures, as indicated. }
\end{figure}

The electronic band structure described by the Hamiltonian (1) consists of four
energy bands,
\begin{equation}
\varepsilon _{{\bf k}n}=\pm \alpha \pm (\hbar ^2v_F^2k^2+\alpha ^2)^{1/2},
\end{equation}
where $n=1 - 4$ is the band index, with $n=1$ ($n=4$) corresponding to the band of the lowest (highest) energy.
Each of the bands is parabolic at small wavevectors, $k\ll \alpha /\hbar v_F$, and has
almost linear dispersion for $k\gg \alpha /\hbar v_F$. Two of these bands ($n=2,3$) touch at $k=0$, while two others ($n=1,4$) are separated by
an energy gap of width equal to $2\alpha$.

We assume the axis $x$ is along the graphene ribbon of length $L$, as shown schematically in Fig.~1.
The ribbon is wide enough to neglect size quantization.
In turn, the length $L$ is smaller than the mean free path $\ell$, $L\ll \ell $, so electronic
transport can be considered as fully ballistic.
For simplicity, we neglect any scattering of electrons inside the ribbon. Additionally,
we assume the graphene ribbon is in contact at the ends with two 2D
electronic reservoirs, which generally have different  temperatures, $T_1$ and $T_2$, as indicated in Fig.1.
Electrons in these reservoirs are
described by the corresponding equilibrium Fermi-Dirac distribution functions,
\begin{eqnarray}
\label{2}
f_{1,2}(\varepsilon _{{\bf k}n})
=[\exp ((\varepsilon _{{\bf k}n}-\mu_{1,2} )/k_BT_{1,2})+1]^{-1},
\end{eqnarray}
where $\mu_1$ and $\mu_2$ are  the  chemical potentials.
Though we are focused mainly on thermal effects, we assume that $\mu_1$ and $\mu_2$  can be different in a general situation.
Thus, the electron system in the ballistic region can be described by the distribution
functions $f^>_1(\varepsilon _{{\bf k}n})$ and $f^<_2(\varepsilon _{{\bf k}n})$
for electrons moving from left to the right and from the right to left, respectively.
In the following we use this distribution to calculate transport and thermoelectric properties of
the graphene ribbon, assuming purely ballistic regime.

\section{Thermally induced charge and heat currents}

Assume equal chemical potentials in the two electronic reservoirs, $\mu_1=\mu_2=\mu$, and different temperatures, $T_1>T_2$.
The former condition will be relaxed only when necessary. Below we calculate charge current due to the temperature difference (gradient),
and also the corresponding electronic contribution to the heat current.

\subsection{Charge current}

The charge current in the ballistic transport regime,  flowing along the axis $x$  due to the
difference $\Delta T=T_1-T_2$ in the reservoir temperatures, can be calculated with the formula
\begin{eqnarray}
\label{3}
j=e\sum _n{\sum _{\bf k}}'
\left< {\bf k}n|\hat{v}_x|{\bf k}n\right>
[f^>_1(\varepsilon _{{\bf k}n})-f^<_2(\varepsilon _{{\bf k}n})],
\end{eqnarray}
where $e$ is the electron charge, $\hat{v}_x=v_F\tau _x$ is the electron velocity operator, and the summation
over the wavevector ${\bf k}$ is restricted to the angles, for which
the $x$-component of electron velocity, $v_{xn}=\left< {\bf k}n|\hat{v}_x|{\bf k}n\right> $,
is positive.

For definiteness, we assume that the chemical potential of electrons is positive, $\mu >0$.
It is clear that due to the electron-hole symmetry, the results for
$\mu <0$ can differ only in sign from those for $\mu <0$. This is because  the electron velocity
$v_{xn}$ is positive for $k_x>0$ in the energy bands with $\varepsilon _{{\bf k}n}>0$ and
negative for the bands with $\varepsilon _{{\bf k}n}<0$.

\begin{figure}
\includegraphics[width=.8\columnwidth]{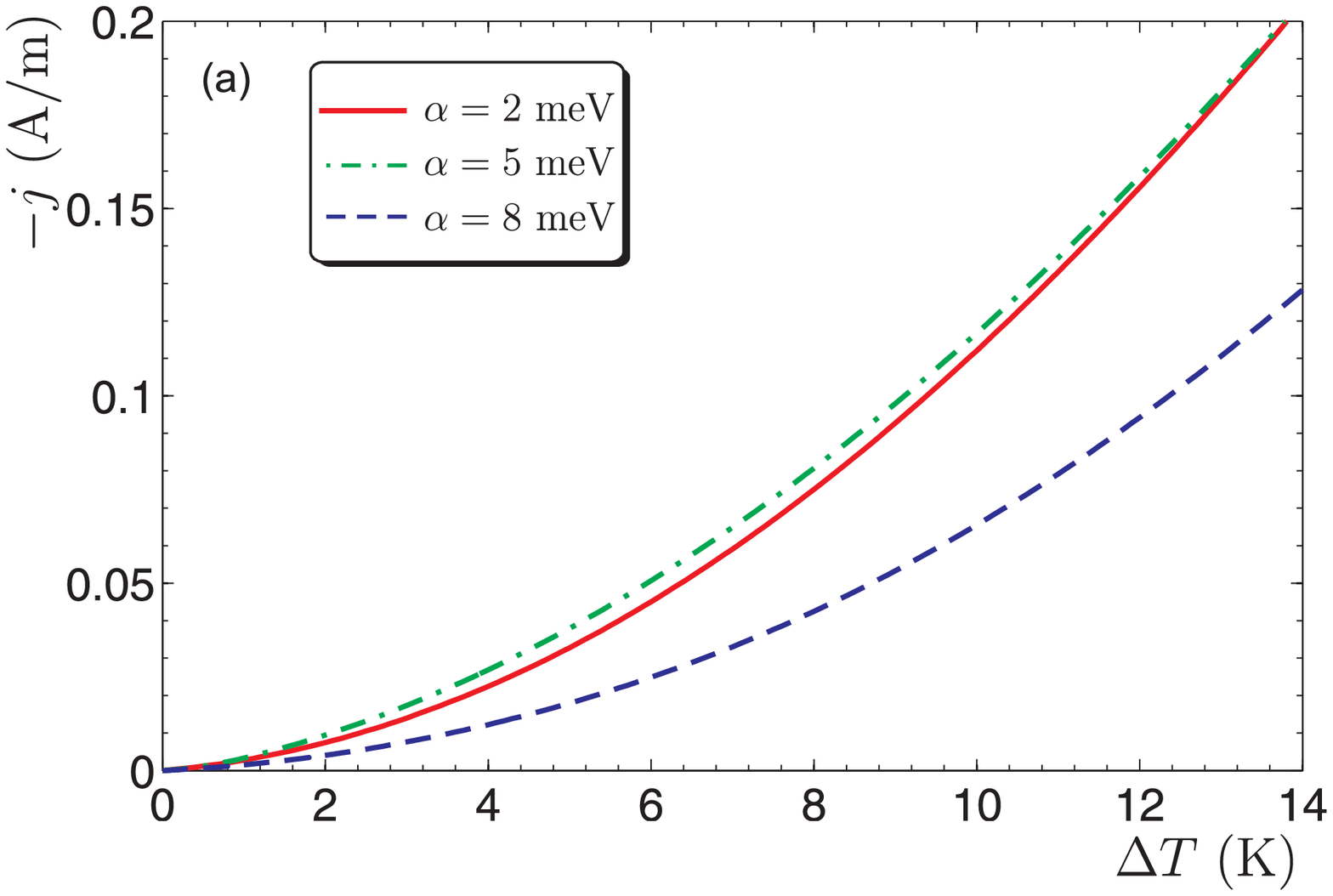}\\
\includegraphics[width=.8\columnwidth]{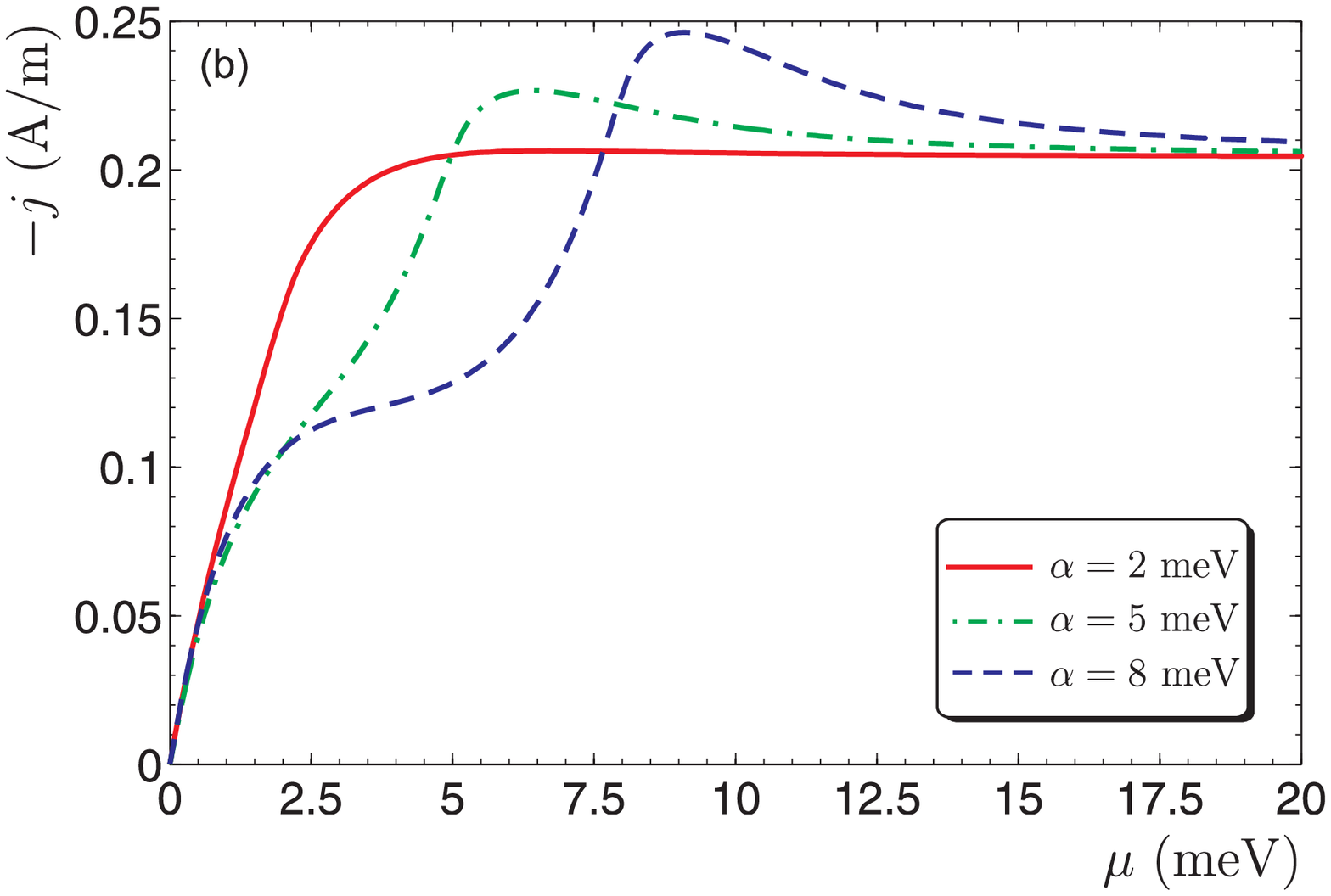}
\caption{Thermoelectric current in the ballistic regime
as a function of $\Delta T$ (a) and chemical potential $\mu $ (b)
for indicated values of the Rashba coupling parameter $\alpha $, and for $\mu=5$~meV (a) and $\Delta T=14$~K (b).}
\end{figure}

The thermoelectric current calculated
as a function of $\Delta T$ is presented in Fig.~2a for
different values of the Rashba coupling constant $\alpha $.
Using Eq.~(3) one can show that the dependence
of current on $\Delta T$ is linear for $\alpha =0$ and
$\Delta T\ll T_1,\, \mu /k_B$, which is related to the
linearity of the density of states in graphene, $\nu (\varepsilon )\sim |\varepsilon |$.
It should be noted that in the case of ordinary 2D electron gas with parabolic energy spectrum,
the thermoelectric current at these conditions is absent since
the corresponding 2D density of states, $\nu _{2D}$, is independent of the electron energy,
and thus the currents due to particles and holes compensate each other, so the net current vanishes.
Thus, the nonzero thermoelectric ballistic current in graphene is related to
the relativistic energy spectrum of this 2D material.
As follows from Fig.2a, the dependence of the thermoelectric current on the Rashba spin-orbit constant
$\alpha $ is non-monotonic, and the current is maximal when $\alpha \simeq \mu $.

\begin{figure}[]
\label{fig4}
\includegraphics[width=.8\columnwidth]{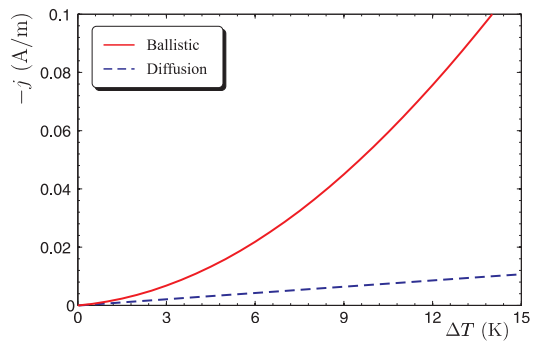}
\caption{Comparison of the thermoelectrically induced current in the ballistic and in the diffusive transport regimes,
calculated for $\mu =5$~meV. In the diffusive regime, the electric current is proportional to the
temperature gradient $\Delta T$.}%
\end{figure}

Variation of the thermoelectric current with the chemical potential $\mu$ is shown in Fig.2b.
First, the current vanishes when the Fermi level is at the particle-hole symmetry point, $\mu =0$.
Second, there is a small peak for larger values of $\alpha $.
Position of this peak corresponds
to the onset of the contribution from another (higher in energy) band, whose bottom band-edge is at the energy
$\varepsilon =\alpha $.
At larger values of $\mu $, the electric current saturates at a  constant value (independent of $\alpha$), since the energy spectrum
is linear in this region, like in graphene without spin-orbit coupling.

To compare the results for ballistic and diffusive junctions, we note that Eq.~(3)
also gives the thermoelectric current in the diffusive regime if we take $\Delta T=L\nabla T$ and
substitute $L$ by the mean free path $\ell $, as it should be to match the ballistic and
diffusive results. Thus, if we keep $\nabla T={\rm const}$ and reduce $\ell $, i.e., if we
go to the diffusive regime by increasing the density of impurities, $\ell <L$, we
decrease the current. The difference between the thermoelectric currents for ballistic and diffusive transport regimes
is shown in Fig.~3. While in the diffusive regime the thermo-current increases linearly with the temperature difference $\Delta T$,
its increase with $\Delta T$ in the ballistic limit is faster and nonlinear.

\begin{figure}
\label{fig5}
\includegraphics[width=0.8\columnwidth]{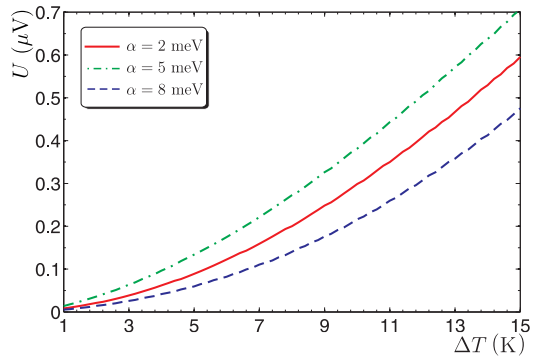}
\caption{Thermoelectric voltage as a function of $\Delta T$, calculated for $\mu=5$~meV and indicated values of the Rashba parameter $\alpha$. }
\end{figure}

One can also calculate the thermopower, or equivalently the thermally induced voltage $U$ between the reservoirs
under the condition of zero charge current, $j=0$. To determine the voltage $U$, one can use
Eq.~(3) with the distribution functions corresponding to different electrochemical potentials, $\mu _1\ne \mu _2$,
as in Eq. 2.
Then, the  voltage $U$ can be  determined from the condition that the thermally induced current is fully
compensated by the field-induced current. The thermoelectric voltage is then given as $U=(\mu_1-\mu _2)/e$.
Results of the corresponding numerical calculation are presented in Fig.~4. As follows from this figure, the thermoelectric voltage depends on
the temperature difference in a nonlinear way, whereas
the dependence on the Rashba parameter is nonmonotonous.

\subsection{Heat current}

\begin{figure}
\label{fig6}
\includegraphics[width=.8\columnwidth]{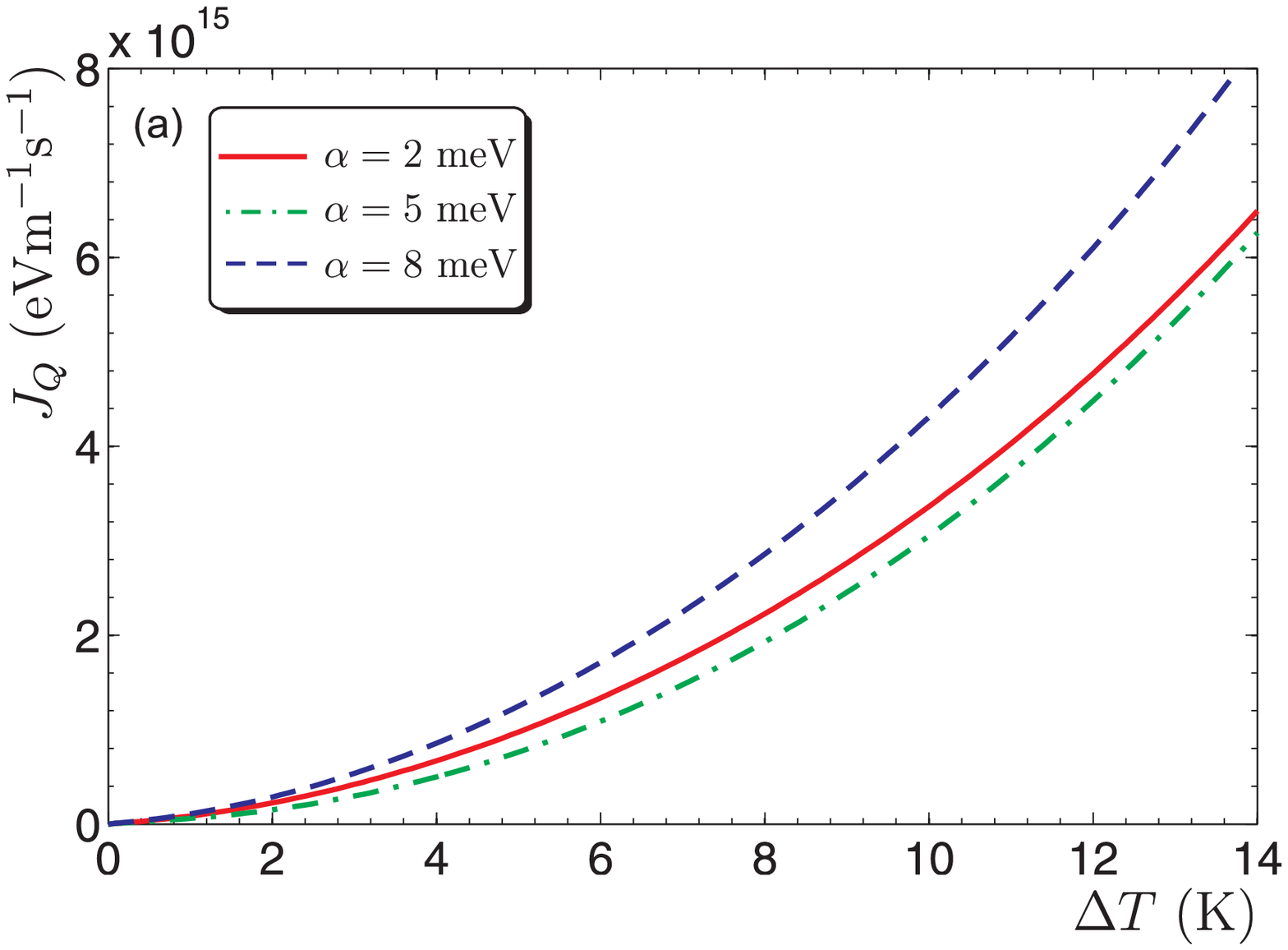}\\
\includegraphics[width=.8\columnwidth]{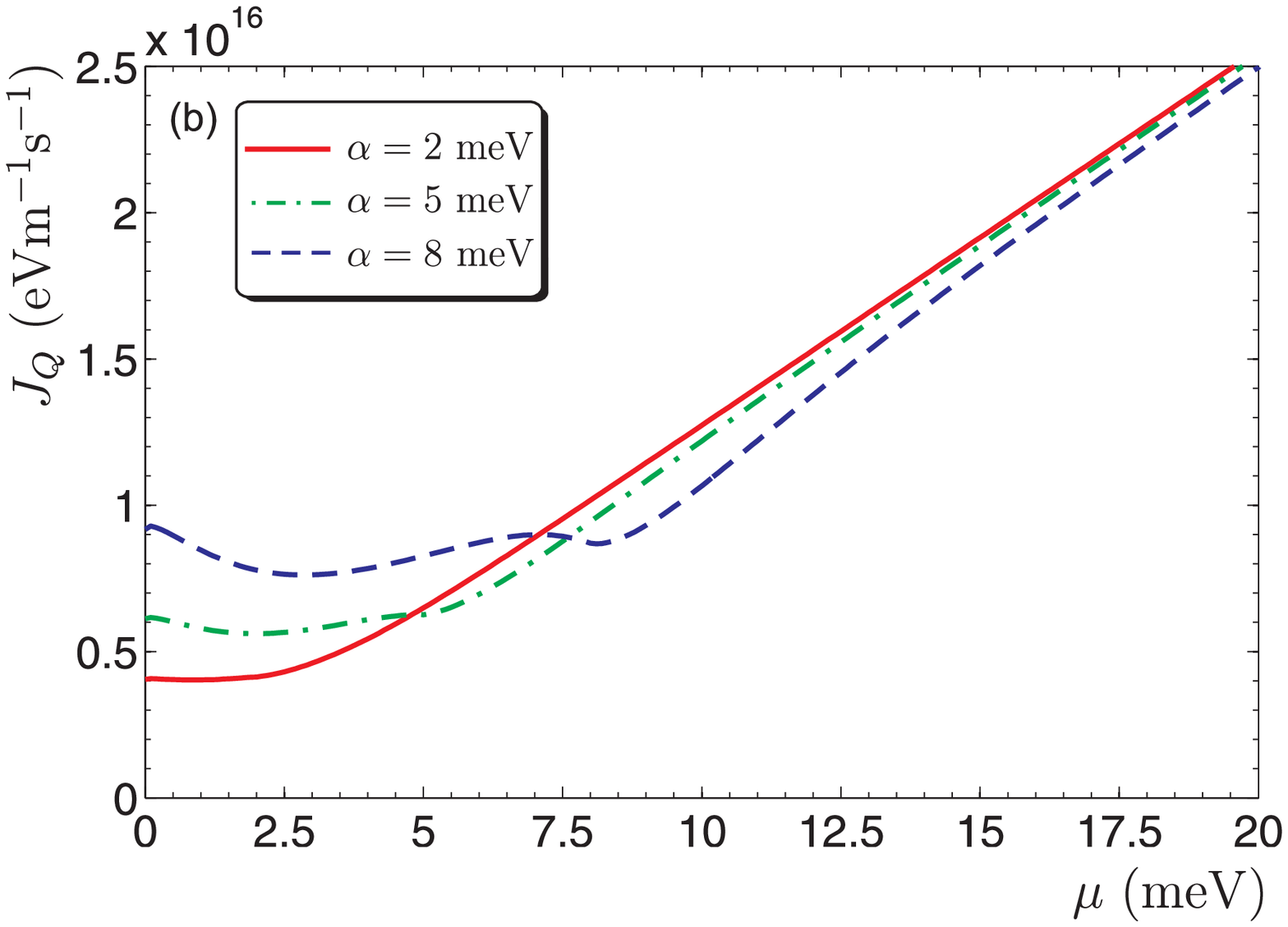}
\caption{Heat flux transmitted by electrons for indicated values of the Rashba parameter $\alpha $, calculated as a function of $\Delta T$
for $\mu =5$~meV (a), and as a function of $\mu$ for $\Delta T=14$~K (b).}
\end{figure}

Employing the same method as that applied above in the calculation of thermoelectric charge current,
one can also calculate the heat current associated with transfer of electrons between the two reservoirs.
The corresponding formula for the heat flux from the left reservoir (of temperature $T_1$) to the right one (of temperature $T_2$)
can be written in the form
\begin{eqnarray}
\label{4}
J_Q=\frac12 \sum _n{\sum _{\bf k}}'
\left< {\bf k}n|\big\{ (H_{\bf k}-\mu),\, \hat{v}_x\big\}|{\bf k}n\right>
\nonumber \\ \times
\big[ f^>_1(\varepsilon _{{\bf k}n})
-f^<_2(\varepsilon _{{\bf k}n})\big] ,
\end{eqnarray}
where $\{\hat{A},\, \hat{B}\}=\hat{A}\hat{B} + \hat{B} \hat{A}$ for any two operators $\hat{A}$ and $\hat{B}$.

Dependence of the heat flux $J_Q$ on the temperature difference $\Delta T$ is presented
in Fig.~5a for different values of the Rashba parameter $\alpha $. Similarly as in the case of charge current, the heat current increases nonlinearly with
 $\Delta T$, and also depends nonmonotonously on the Rashba parameter $\alpha $, compare Fig.2a and Fig.5a.

 Figure 5b in, turn, shows the dependence of the heat current  $J_Q$ on the chemical potential $\mu$. As follows from this figure, the dependence on $\mu$ is linear at large values of $\mu $, contrary to the behavior of charge current which saturates at large $\mu$ (see Fig.2b). This difference appears because the contribution to heat current from a transferred electron depends on its  energy, $\varepsilon \sim \mu $, while the corresponding contribution to charge current is independent on this energy.

\section{Thermally induced spin polarization and spin current}

As in the preceding section, we assume equal chemical potentials in the two electronic reservoirs, $\mu_1=\mu_2=\mu$.
Below we calculate spin polarization of electrons and spin current, both induced by a temperature difference $\Delta T$.

\subsection{Spin polarization}

\begin{figure}[t]
\label{fig8}
\includegraphics[width=.8\columnwidth]{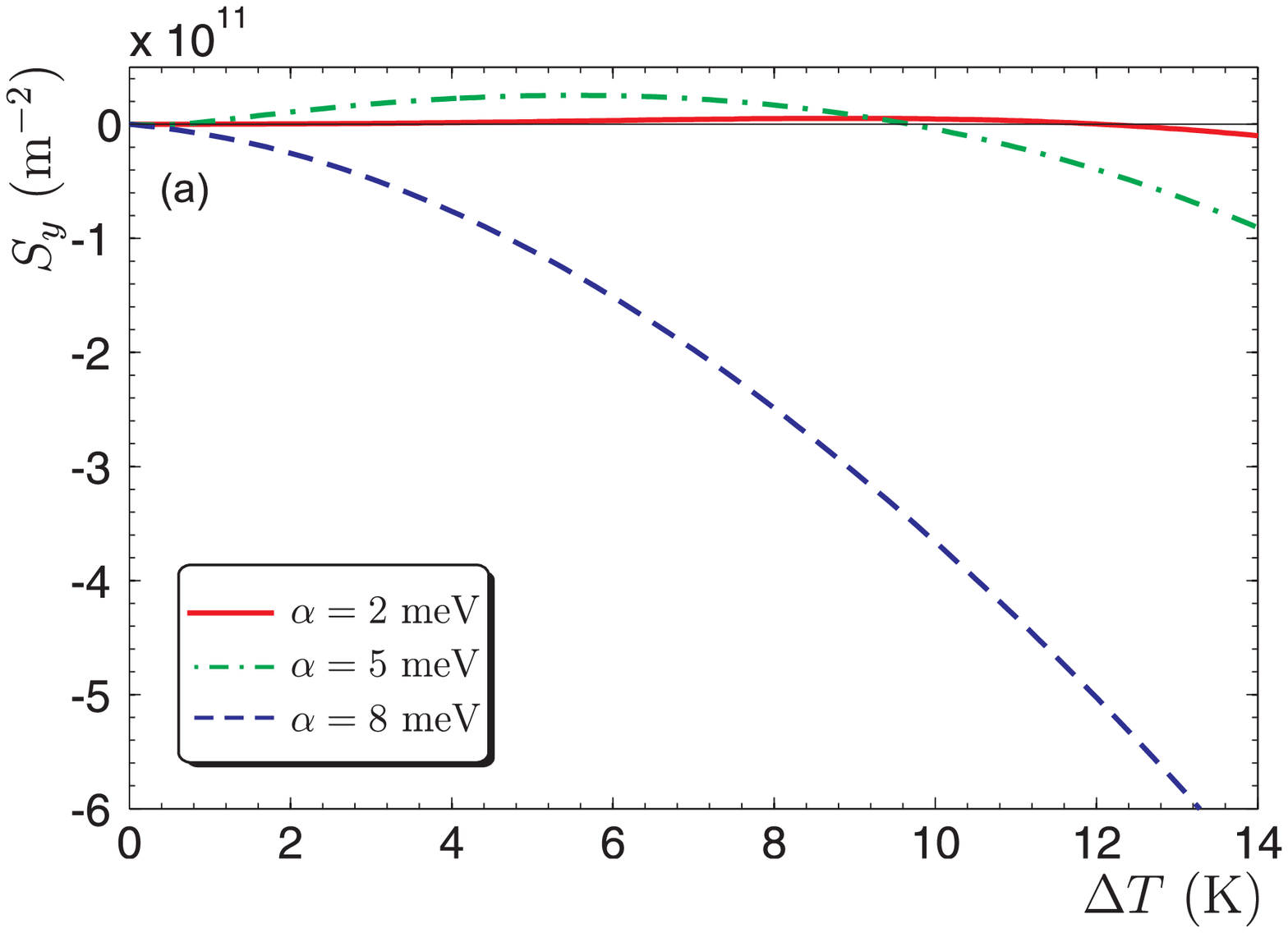}\\
\includegraphics[width=.8\columnwidth]{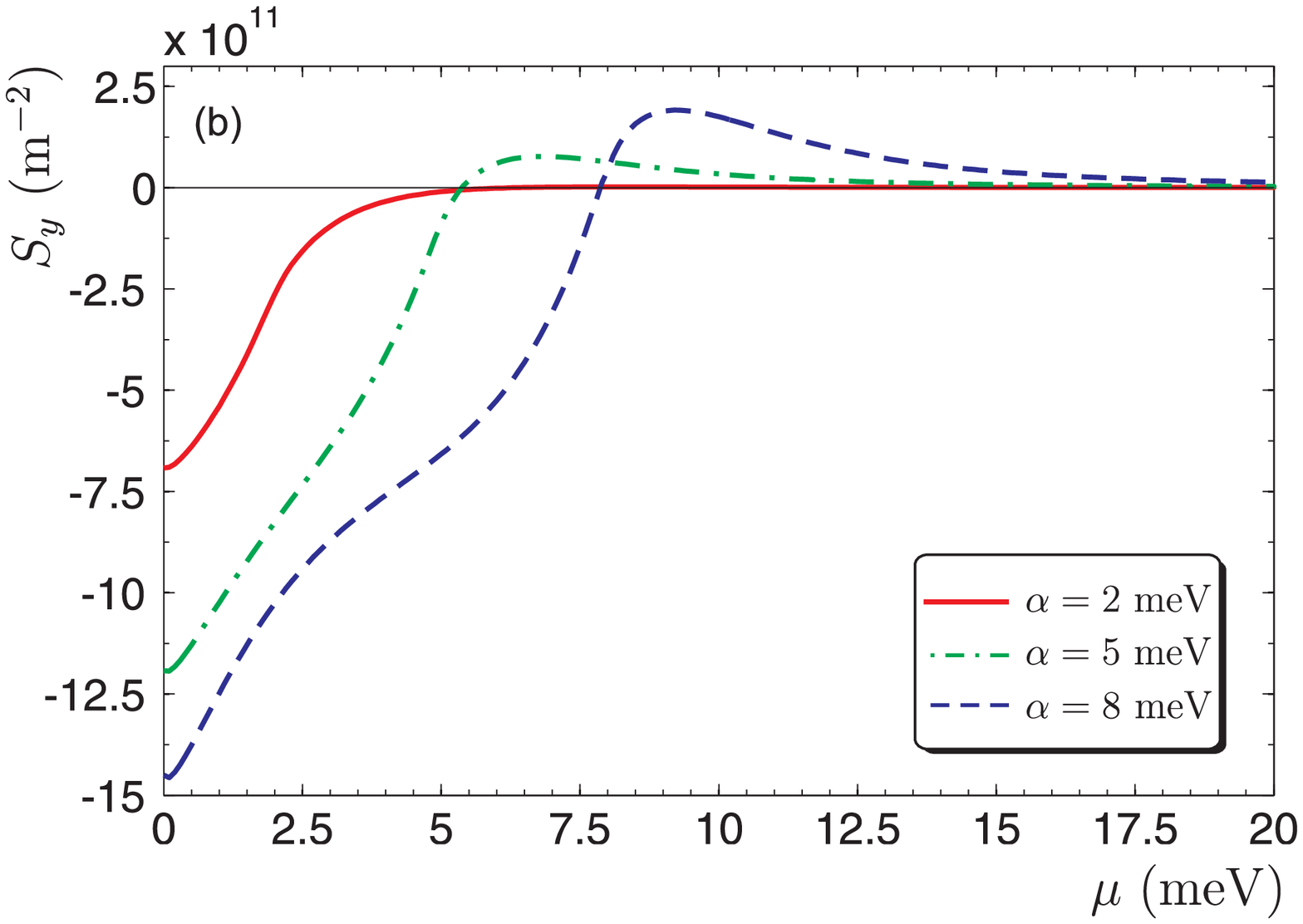}
\caption{Thermoelectrically induced spin polarization as a function of $\Delta T$ (a) and as a function of $\mu$ (b), calculated for indicated values of the Rashba parameter and  for $\mu =5$~meV. (a) and $\Delta T=14$~K (b).}%
\end{figure}

It is already well known that spin-orbit interaction in the presence of either electric field or temperature gradient can induce spin polarization of conduction electrons.  This effect has been  studied theoretically for the usual 2D electron gas with parabolic energy spectrum
and Rashba spin-orbit coupling,~\cite{Edelstein,Aronov,Dyrdal13} as well as in graphene.\cite{Dyrdal14}
In the ballistic junction considered in this paper, the spin density can be calculated using the formula
\begin{eqnarray}
\label{5}
S_\alpha =\sum _n{\sum _{\bf k}}' \left< {\bf k}n|\sigma _\alpha |{\bf k}n\right>
\big[ f^>_1(\varepsilon _{{\bf k}n})-f^<_2(\varepsilon _{{\bf k}n})\big] .
\end{eqnarray}

The corresponding numerical results are presented in Fig.~6, where the thermally-induced spin polarization $S_y$ is shown as a function of $\Delta T$ (Fig.6a)
and as a function of $\mu $ (Fig.6b). The induced spin polarization is in the graphene plane, and is normal to the temperature gradient, similarly as in the case of 2D electron gas.\cite{Dyrdal13}
Physical mechanism of the spin polarization in graphene with Rashba spin-orbit coupling
is also similar to the mechanism of spin polarization in 2D electron gas.
Indeed, there is a nonzero spin polarization of an electron in the eigenstate
$\left| {\bf k}n\right> $, which is perpendicular to the wavevector ${\bf k}$. The
magnitude of this spin polarization is small for $k\ll \alpha /\hbar v_F$ and
is equal to its maximum value equal to $\hbar /2$ for $k\gg \alpha /\hbar v_F$.
\cite{Rashba09} A nonzero spin
polarization $S_y$ appears due to the imbalance of the distribution of electrons with
$k_x>0$ and $k_x<0$.\cite{Dyrdal14}

\subsection{Spin currents}

\begin{figure}[b]
\label{fig10}
\includegraphics[width=0.8\columnwidth]{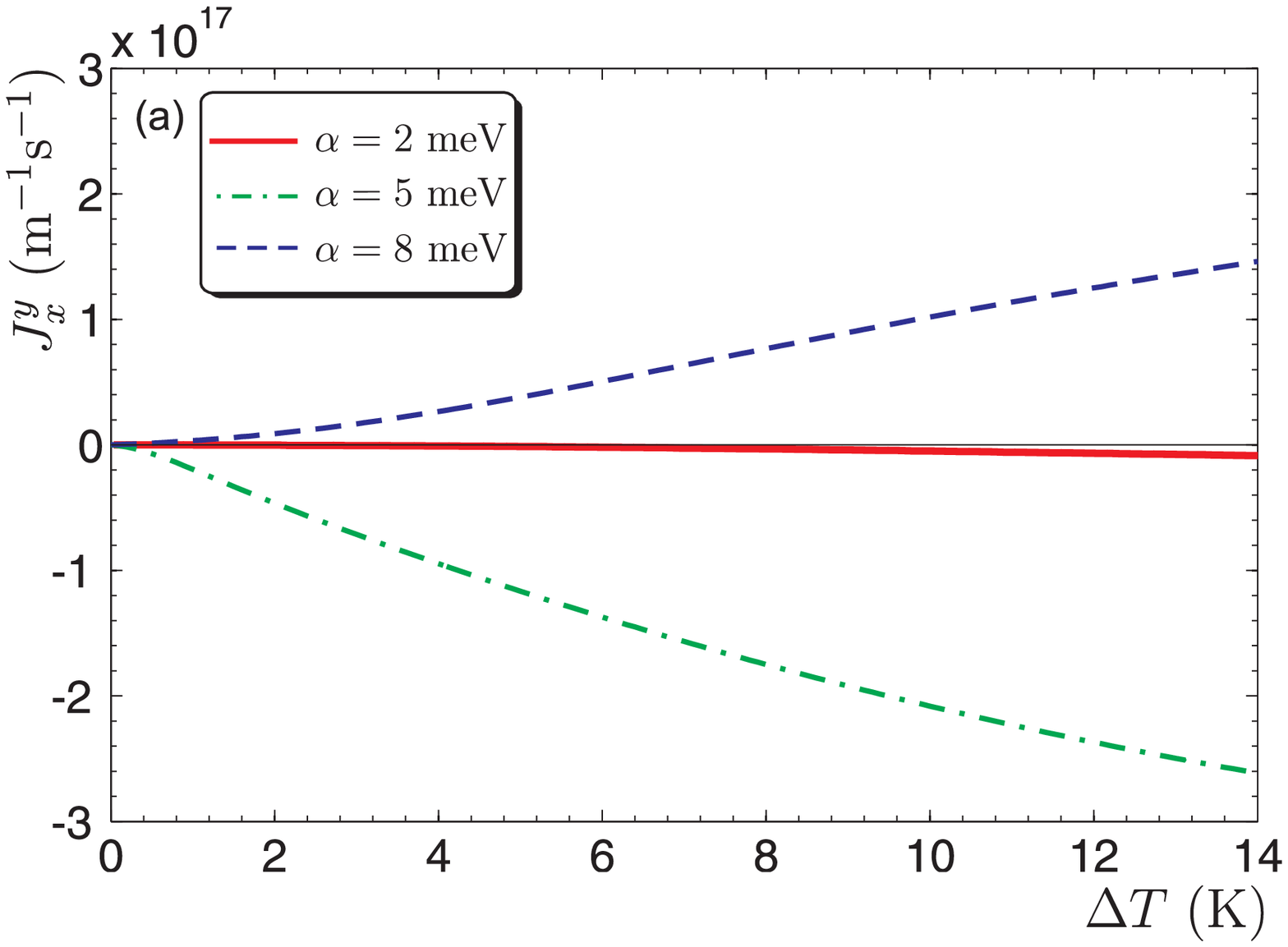}\\
\includegraphics[width=0.8\columnwidth]{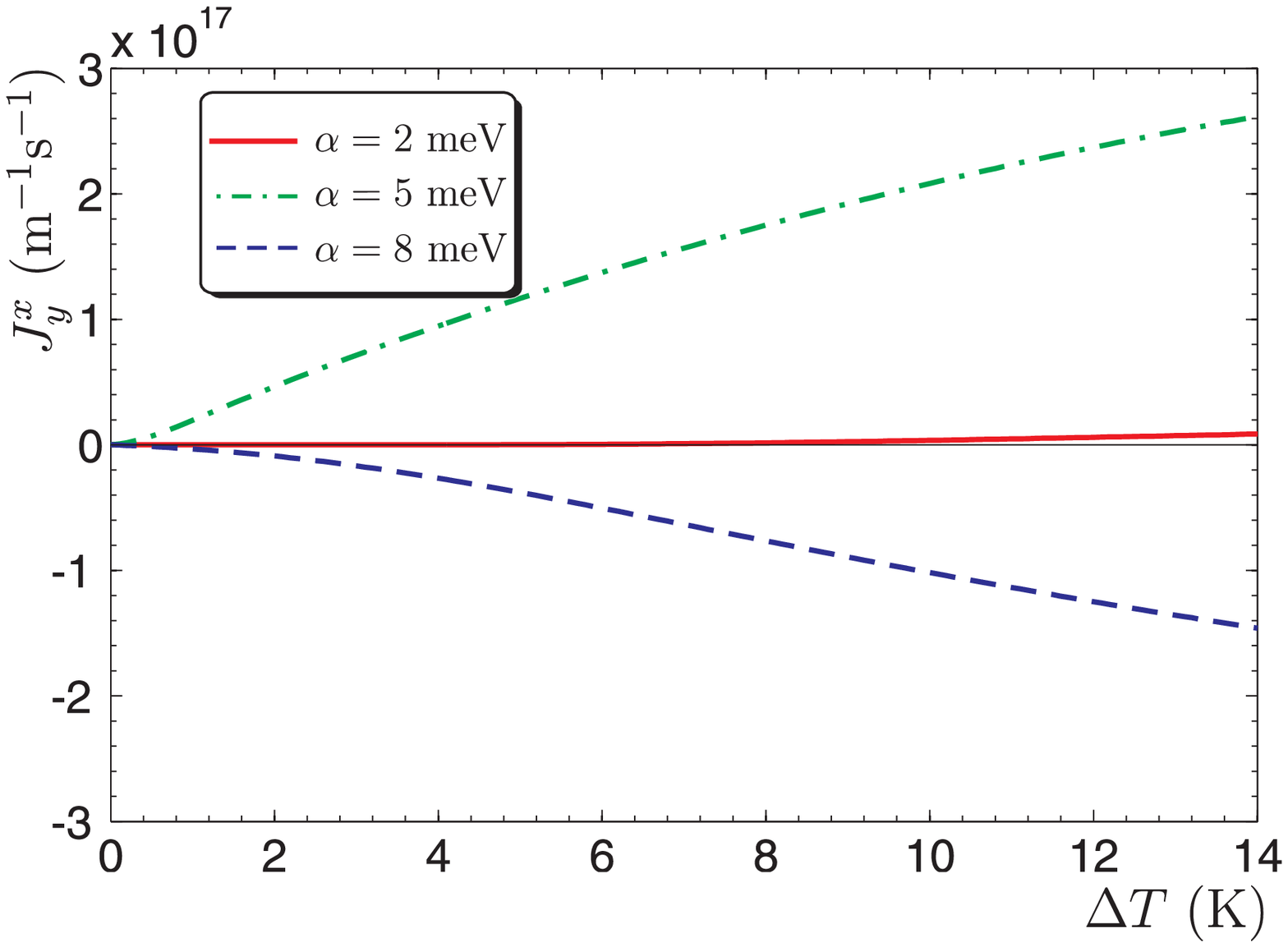}
\caption{Thermally induced spin currents $J_x^y$ (a) and $J_y^x$ for different values of $\alpha $, calculated as a function of $\Delta T$
for $\mu =5$~meV. (a).  }
\end{figure}
\begin{figure}[]
\label{fig11}
\includegraphics[width=0.8\columnwidth]{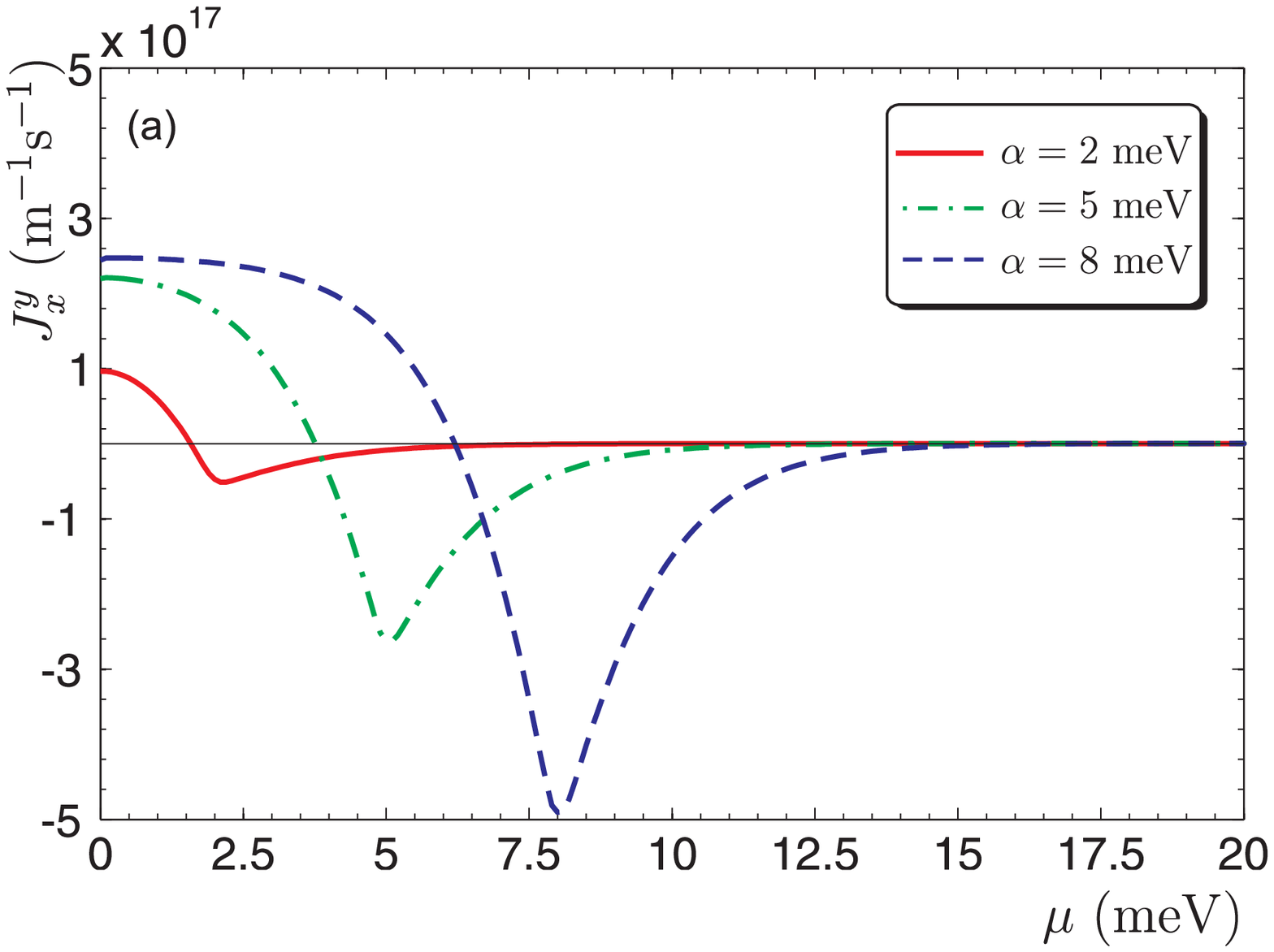}\\
\includegraphics[width=0.8\columnwidth]{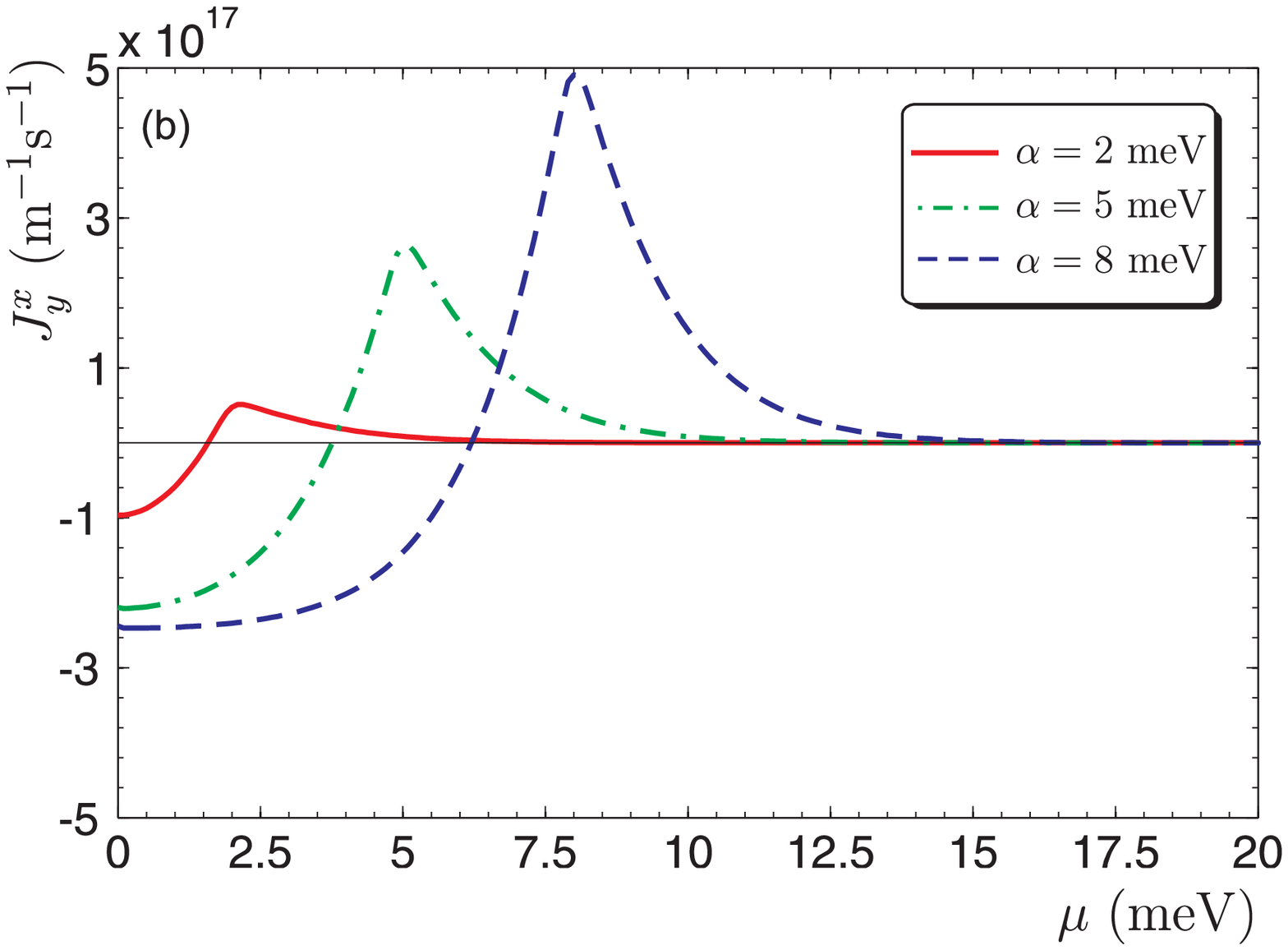}
\caption{Thermally induced spin currents $J_x^y$ (a) and $J_y^x$ for different values of $\alpha $,
 calculated as a function of $\mu $ for $\Delta T= 14$~K.}
\end{figure}

The temperature difference between the electron reservoirs can also generate a spin current $J_x^y$ flowing parallel to the temperature  gradient as well  the spin current $J_y^x$ flowing  perpendicularly to the gradient. Here the upper index indicated the spin component associated with the spin current, while the lower index indicated the orientation of the spin current flow.  Both components of the spin current can be calculated  using the same method as in the case of charge current. The relevant formula takes now the form
\begin{eqnarray}
\label{6}
J^\alpha _i=\frac12 \sum _n{\sum _{\bf k}}'
\left< {\bf k}n|\big\{ \sigma _\alpha , \hat{v}_i \big\} |{\bf k}n\right>
\big[ f^>_1(\varepsilon _{{\bf k}n})-f^<_2(\varepsilon _{{\bf k}n})\big] .
\end{eqnarray}

The numerical results for both $J^y_x$ and $J^x_y$ calculated as a function of the temperature difference $\Delta T$ are presented in Fig.~7 and as a function of the chemical potential $\mu$ in Fig.~8. The mechanism of a nonzero
component $J^y_x$ is related to the spin polarization of electrons due to the temperature gradient and Rashba spin-orbit interaction, as calculated and discussed above. These spin polarized electrons are transferred between the two electron reservoirs, giving rise to the spin current $J^y_x$.
In turn, the other spin current component, $J^x_y$, corresponds to the thermally induced
spin Hall effect, called also the spin Nernst effect. This effect consists in a spin current generation by a temperature gradient. 
The induced current flows then perpendicularly to the temperature gradient.  

\section{Summary}

We have analyzed thermoelectric and thermospin effects in a ballistic graphene ribbon attached to two electronic reservoirs of different temperatures.
The graphene ribbon was assumed to be deposited an a substrate that generated a strong spin-orbit coupling of Rashba type. We have calculated not only the thermally induced charge current between the two reservoirs, and the associated thermoelectric voltage, but also thermally induced spin polarization and spin current.
Numerical results on the charge current show that the current in the ballistic regime is significantly larger than in the diffusive one.

The spin current, in turn,  is shown to have two components. One of them is  related to the thermally induced spin polarization of electrons transferred from one reservoir to the other, while the other one reveals the spin Nernst effect, i.e. the thermally induced spin Hall effect. We have also calculated the heat transferred by ballistic electrons from the reservoir of higher temperature to the reservoir of lower temperature.

\begin{acknowledgments}
This work was supported by the National Science Center in Poland as research projects Nos. DEC-2012/06/M/ST3/00042 (MI), and DEC-2012/04/A/ST3/00372 (VKD and JB). The work of (MI) is also supported by the projects POIG.01.04.00-18-101/12 and UDA-RPPK.01.03.00-18-025/13-00.
\end{acknowledgments}


\begin{thebibliography}{99}

\bibitem{castro09}
A. H. Castro Neto, F. Guinea, N. M. R. Peres, K. S. Novoselov, and A. K. Geim,
Rev. Mod. Phys., \textbf{81}, 109 (2009).

\bibitem{sonin09}
E. B. Sonin, \prb {\bf 79}, 195438 (2009).

\bibitem{katsnelson06}
M. I. Katsnelson, K. S. Novoselov, and A. K. Geim, Nature Physics {\bf 2}, 620 (2006).

\bibitem{lee15}
G. H. Lee,	S. Kim, S. H. Jhi, and H. J. Lee, Nature Communications, \textbf{6}, 6181 (2015).

\bibitem{borunda13}
M. F. Borunda, H. Hennig, and E. J. Heller, \prb {\bf 88}, 125415 (2013).

\bibitem{grushina13}
 A. L. Grushina, D. K. Ki, and A. F. Morpurgo, Appl. Phys. Lett., {\bf 102}, 223102 (2013).

\bibitem{nguyen14}
N.T. T. Nguyen, D. Q. To, and V. L. Nguyen, J. Phys. Condens. Matter {\bf 26}, 015301 (2014).

\bibitem{titov06}
M. Titov, and C. W. J. Beenakker, \prb (R) {\bf 74}, 041401 (2006).

\bibitem{Yamakage2011}
A. Yamakage, K. I. Imura, J. Cayssol, Y. Kuramoto, Phys. Rev. B, 83, 125401 (2011).

\bibitem{Rataj13}
M. Rataj and J. Barna\'s, Phys. Status Solidi: Rapid Res. Lett.
7, 997 (2013).

\bibitem{balandin2008}
A. A. Balandin, S. Ghosh, W. Bao, I. Calizo, D. Teweldebrahn, F. Miao, and C. Lau,
Nano Lett. \textbf{8}, 902 (2008).

\bibitem{Wierzbowska1}
M. Wierzbowska, A. Dominiak, and G. Pizzi, 2D Materials {\bf 1}, 035002 (2014).

\bibitem{Wierzbowska2}
 M. Wierzbowska and A. Dominiak, Carbon {\bf 80}, 255 (2014).

\bibitem{Wierzbicki13}
M. Wierzbicki, R. Swirkowicz, and J. Barna\'s,  Phys. Rev. B {\bf 88}, 235434 (2013).

\bibitem{foster09}
M. S. Foster and I. L. Aleiner, \prb 79, 085415 (2009).

\bibitem{wei09}
P. Wei, W. Bao, Y. Pu, C. N. Lau, and J. Shi, Phys. Rev. Lett. \textbf{102}, 166808 (2009).

\bibitem{yan09}
X. Z. Yan, Y. Romiah, and C. S. Ting, \prb {\bf 80}, 165423 (2009).

\bibitem{zuev09}
Y. M. Zuev, W. Chang, and P. Kim, Phys. Rev. Lett. \textbf{102}, 096807 (2009).

\bibitem{ugarte11}
V. Ugarte, V. Aji, and C. M. Varma, \prb {\bf 84}, 165429 (2011).

\bibitem{wang11}
D. Wang and J. Shi, Phys. Rev. B {\textbf{83}}, 113403 (2011).

\bibitem{sharapov12}
S. G. Sharapov and A. A. Varlamov, Phys. Rev. B, \textbf{86}, 035430 (2012).

\bibitem{alomar14}
M. I. Alomar and D. S\'anchez, \prb {\bf 89}, 115422 (2014).

\bibitem{Seol10}
J. H. Seol, I. Jo, A. L. Moore, L. Lindsay, Z. H. Aitken, M. T. Pettes, X. Li, Z. Yao, R. Huang,
D. Broido, N. Mingo, R. S. Ruoff, L. Shi, Science \textbf{328}, 213 (2010).

\bibitem{Dragoman}
D. Dragoman and M. Dragoman, Appl. Phys. Lett. \textbf{91}, 203116 (2007).

\bibitem{stauberPRB07}
T. Stauber, N. M. R. Peres, and F. Guinea \prb {\bf 76}, 205423, (2007).

\bibitem{Inglot15}
M. Inglot, A. Dyrdal, V. K. Dugaev, and J. Barna\'s, Phys. Rev. B {\bf 91}, 115410 (2015).

\bibitem{Xu2014}
Y. Xu, Z. Li, and W. Duan, Small {\bf 10}, 2182 (2014).

\bibitem{Zheng09}
X. H. Zheng, G. R. Zhang, Z. Zeng, V. M. Garcia-Suarez, and C. J. Lambert,
Phys. Rev. B 80, 075413 (2009).

\bibitem{Zhang10}
Y.-T. Zhang, Q. M. Li, Y. C. Li, Y. Y. Zhang, and F. Zhai, J. Phys.:
Condens. Matter 22, 315304 (2010).

\bibitem{aksamijaPRB14}
Z. Aksamija, and I. Knezevic, \prb {\bf 90}, {035419}, (2014).

\bibitem{sevincli10}
H. Sevincli and G. Cuniberti, \prb 81, 113401 (2010).

\bibitem{huang11}
W. Huang, J. S. Wang, G. Liang, \prb {\bf 84}, 045410 {2011}.

\bibitem{chang12}
P. H. Chang and B. Nikoli\'c, Phys. Rev. B {\bf 86}, 041406(R) (2012).

\bibitem{liang12}
L. Liang, E. C. Cruz-Silva E. C. Girao, and V. Muenier, \prb {\bf 86}, 115438 (2012).

\bibitem{chenPRB14}
W. Chen and A. A. Clark, \prb {\bf 86}, {125443} (2012).

\bibitem{lindsayPRB14}
L. Lindsay, Wu Li, J. Carrete, N. Mingo, D. A. Broido, and T. L. Reinecke,
\prb {\bf 89}, 155426, (2014).

\bibitem{jaffres14}
H. Jaffr\`es, Physics {\bf 7}, 123 (2014).

\bibitem{oltscher14}
M. Oltscher, M. Ciorga, M. Utz, D. Schuh, D. Bougeard, and D. Weiss.
\prl {\bf 113}, 236602 (2014).

\bibitem{mayorov11}
A. S. Mayorov, R. V. Gorbachev, S. V. Morozov, L. Britnell, R. Jalil,
L. A. Ponomarenko, P. Blake, K. S. Novoselov, K. Watanabe, T. Taniguchi,
and A. K. Geim, Nano Letters {\bf 11}, 2396 (2011).

\bibitem{baringhaus14}
J. Baringhaus, M. Ruan,	F. Edler,	A. Tejeda,	M. Sicot,	A. Taleb-Ibrahimi,	A. P. Li,	Z. Jiang,
E. H. Conrad, C. Berger,	C. Tegenkamp	and W. A. de Heer, Nature {\bf 506}, 349 (2014).

\bibitem{Dyrdal14}
A. Dyrdal, J. Barna\'s, and V. K. Dugaev, Phys. Rev. B {\bf 89}, 075422 (2014).

\bibitem{Edelstein}
V. M. Edelstein, Sol. State Communs. {\bf 73}, 233 (1990).

\bibitem{Aronov}
A. G. Aronov and Y. B. Lynda-Geller, JETP Lett. {\bf 50}, 431 (1989).

\bibitem{Dyrdal13}
A. Dyrdal, M. Inglot, V. K. Dugaev, and J. Barna\'s, Phys. Rev. B {\bf 87}, 245309 (2013).

\bibitem{Rashba09}
E. I. Rashba, Phys. Rev. B {\bf 79}, 161409 (2009).



\end{thebibliography}

\end{document}